\documentclass{article}
	\usepackage{fullpage, graphicx}
	\author{Sujit S. Datta$^{1}$, Douglas R. Strachan$^{1, 2}$\thanks{E-mail: \texttt{drstrach@sas.upenn.edu}}, Samuel M. Khamis$^{1}$, A. T. Charlie Johnson$^{1}$\thanks{E-mail: \texttt{cjohnson@physics.upenn.edu}}\\ \\ {\normalsize 1 -- Department of Physics and Astronomy,}\\ {\normalsize 2 -- Department of Materials Science and Engineering},\\{\normalsize University of Pennsylvania, Philadelphia PA 19104}}	
	\date{{\normalsize Received: February 27, 2008; Revised Manuscript Received June 9, 2008}}
	\title{Crystallographic Etching of Few-Layer Graphene}
	\begin{document}
	\maketitle
\begin{abstract}
We demonstrate a method by which few-layer graphene samples can be etched along crystallographic axes by thermally activated metallic nanoparticles. The technique results in long ($>$1 $\mu$m) crystallographic edges etched through to the insulating substrate, making the process potentially useful for atomically precise graphene device fabrication. This advance could enable atomically precise construction of integrated circuits from single graphene sheets with a wide range of technological applications.  \end{abstract}
\begin{center}
\line(1,0){300}\\ 
\end{center}
Due to its remarkable electronic properties, few layer graphene is emerging as a promising new material for use in a vast array of postsilicon nanoelectronic devices incorporating quantum size effects.$^{1, 2}$ Of particular interest would be the construction of atomically precise graphene nanoribbons, in which charge carriers are confined in the lateral dimension whereby the electronic properties are controlled by the width and specific crystallographic orientation of the ribbon.$^{3-14}$ Such structures hold enormous promise as nanoscale devices similar to those recently developed using carbon nanotubes$^{2, 11, 15}$ with the added advantage that grapheneÕs two-dimensionality lends itself to existing device architectures based on planar geometries.\\

However, these structures have so far been impossible to achieve because of the rough noncrystalline edges of the graphene that result from current state-of-the-art nanolithography techniques.$^{2, 16, 17}$ These rough edges are thought to be the crucial limiting factor to attaining useful performance and on/off current ratios from nanoscale graphene devices.$^{13, 18, 19}$\\

As a step toward band gap engineering of this material, we have developed a means by which few-layer graphene (FLG) samples can be etched along crystallographic axes by thermally activated metallic nanoparticles.$^{20}$ The technique results in long ($>$1 $\mu$m) trenches commensurate with the crystal lattice that are etched through to the supporting insulating substrate, making the process potentially useful for atomically precise graphene device fabrication, as well as indicating a possible method by which entire circuits could be carved out from single graphene sheets.$^{1}$\\

Our initial samples (before etching) consisted of pristine few-layer graphene sheets transferred onto highly doped Si substrates with 300 nm thermally grown SiO$_{2}$ by mechanical exfoliation under ambient conditions, similar to the technique described in ref 21. Flakes of few-layer graphene are identified using optical microscopy. Height imaging of our samples is done using a Veeco Dimension 3100 atomic force microscope (AFM) operating in intermittent contact mode, with Si$_{3}$N$_{4}$-coated tips (NSC15, Mikromasch) of curvature radius $\leq$ 20 nm. Samples are also characterized using a JEOL JSM6400 scanning electron microscope (with a LaB$_{6}$ filament).\\

The FLG on SiO$_{2}$/Si substrates are uniformly spin-coated with $\sim$15 mL solution of 50 mg/L Fe(NO$_{3}$)$_{3}\cdot$9H$_{2}$O in isopropyl alcohol (HPLC grade). The samples are then transferred to a furnace and heated in hydrogen and argon gas coflow (320 sccm/600 sccm, respectively) at 900$^\circ$C for 45 min. At these temperatures, the Fe accumulates to form small $\sim$15 nm diameter nanoparticles which can diffuse along the surface of the SiO$_{2}$ and graphene.$^{22}$ As the nanoparticles diffuse over the surface of the SiO$_{2}$ and graphene at these elevated temperatures, they etch away the FLG sheets.\\

Figure 1 shows AFM and SEM images of FLG samples after this etching procedure. Crucially, we observe long and straight ($>$1 $\mu$m) nanotrenches in the FLG along which graphene has been removed. The majority of these trenches are etched down to the insulating substrate. This is evident through contrast variations in SEM images and through detailed AFM height analysis of the nanotrenches (Figure 1). We find through cross-section analysis (as in Figure 2) that the depths of the trenches correspond to the difference in height between the bare substrate and the FLG flake. This indicates that the trench forming process results in tracks that are etched down to the insulating substrate. AFM imaging shows that the graphene surface has roughness on the order of 0.5 nm, which is the roughness of the substrate and is consistent with our and othersÕ measurements of the intrinsic roughness of FLG on SiO$_{2}$:$^{23}$ this indicates that the graphene surface is largely unspoiled by the etching procedure. AFM analysis also reveals that the widths of the trenches are on the order of tens of nanometers and frequently less than 20 nm. In the AFM images, some narrow trenches do not appear to extend down to the underlying substrate which can be due to the fact that they are not fully resolved when they are narrower than the $\sim$20 nm wide AFM tip.\\

Measurements of the lengths and orientations of etched FLG trenches reveal striking correlations with the graphene lattice. Figure 3 shows a histogram of total etched trench length versus angle for the FLG sample of Figure 2A. Typically, particle tracks travel predominantly along a single direction (defined as 0$^\circ$ in Figure 3) with other preferred directions at 60$^\circ$ relative to this. Slightly less preferred directions at 30$^\circ$ intervals are also observed. It is currently not clear why a single etching direction is preferred. In addition, we note that no significant correlation was found between the orientation of the etched nanotrenches and the direction of gas flow. The existence of these track directions spaced at 30$^\circ$ intervals gives strong support to the notion that the etched trenches are commensurate with the graphene honeycomb lattice; thus, our etching procedure could be potentially useful in constructing device edges that are oriented with the underlying crystal lattice of the graphene. This is in contrast to conventional nanolithography etching of FLG samples, which do not yield structures that are commensurate with the graphene lattice. While investigations of the atomic-scale details of our etched graphene edges are still underway, the observed long (on the order of a micrometer) trenches suggest that the edges have long-range order. In addition to other structures, we have used this single-particle etching technique to fabricate graphene nanoribbons of widths measured using AFM as small as $\sim$15 nm and lengths on the order of micrometers (e.g., inset to Figure 1A).\\

FLG etching with metallic nanoparticles likely occurs by a hydrogenation mechanism similar to that studied by numerous researchers several decades ago for bulk carbon allotropes (e.g., ref 20). This reaction, catalyzed by the Fe nanoparticles, is given by
\begin{eqnarray}
\mbox{C(s)} + 2\mbox{H}_{2}\mbox{(g)}\rightarrow\mbox{CH}_{4}\mbox{(g)}
\end{eqnarray}
in which the graphene acts as the carbon source C(s). Further evidence that this reaction is the source of the crystallographic etching is found from experiments performed in hydrogen-free argon gas flow which do not result in nanotrench formation.$^{24}$\\

As the Fe particles diffuse on the SiO$_{2}$ surface, they react with active carbon atoms at the graphene edges. This initiates nanotrench formation, and the moving Fe nanoparticles remove carbon atoms in the graphene along tracks aligned with the underlying crystal lattice, forming the observed nanotrenches (Figure 4). Indeed, we observe that the vast majority of etched trenches commence from the edges of the FLG sheets, with an Fe nanoparticle at the end of each trench. Furthermore, isolated nanoparticles are sometimes observed on top of FLG flakes without any etch track, indicating that carbon atoms at FLG edges (versus those on the FLG surface) are reactive and more likely to initiate nanotrench formation.\\

The fact that the etching occurs as the nanoparticles move over an amorphous SiO$_{2}$ surface indicates that the crystallographic orientation of the trenches are determined from the interfacial interactions of the nanoparticle with the edge of the FLG. This crystallographic etching could be due to the favorable adhesion and wetting of the Fe nanoparticles to the graphene edge along specific crystallographic directions.$^{25}$ Another potential mechanism for our crystallographic etching technique is crystallographic dependence of the Fe-graphene reactivity, for example, due to lowered activation energy of the reaction given in eq 1 along specific directions commensurate with the FLG lattice.$^{20, 26}$\\

In conclusion, we have demonstrated a technique by which few-layer graphene can be etched along crystallographic directions down to the underlying substrate, a potentially useful technique for attaining further progress in the development of graphene nanoelectronic devices. This technique relies on the use of individual metal nanoparticles as catalysts to ``carve out" long, straight nanotrenches with crystallographically oriented edges in FLG samples. The future development of techniques whereby the motion of the nanoparticles can be controlled will lead to possible exciting new graphene device architectures. This could enable the construction of integrated transistors, circuits, chemical sensors, and spin valves from single graphene sheets with many far-reaching potential applications.$^{1}$\\ \\

\begin{center}\noindent{\bf Acknowledgment}\end{center}
This work was supported by the Nano/Bio Interface Center through the National Science Foundation NSEC DMR-0425780 and the JSTO DTRA, the Army Research Office Grant W911NF-06-1-0462, and the Intelligence Community Postdoctoral Fellowship Program.\\
\newpage
\begin{center}\noindent{\bf References}\end{center}
\begin{enumerate}
\item Geim, A. K.; Novoselov, K. S. {\it Nat. Mater.} {\bf 2007}, 6, 183Ð191.   
\item Avouris, P.; Chen, Z.; Perebeinos, V. {\it Nat. Nanotechnol.} {\bf 2007}, 2, 605Ð615. 
\item Ezawa, M. {\it Phys. Rev. B} {\bf 2006}, 73, Art. No. 045432. 
\item Son, Y.-W.; Cohen, M. L.; Louie, S. G. {\it Phys. Rev. Lett.} {\bf 2006}, 97, Art. No. 216803.
\item Brey, L.; Fertig, H. A. {\it Phys. Rev. B} {\bf 2006}, 73, Art. No. 235411.
\item Shemella, P.; Zhang, Y.; Mailman, M.; Ajayan, P. M.; Nayak, S. K. {\it Appl. Phys. Lett.} {\bf 2007}, 91, Art. No. 042101. 
\item Son, Y.-W.; Cohen, M. L.; Louie, S. G. {\it Nature} {\bf 2006}, 444, 347Ð349.  
\item Nakada, K.; Fujita, M.; Dresselhaus, G.; Dresselhaus, M. S. {\it Phys. Rev. B} {\bf 1996}, 54, 17954Ð17961. 
\item Wakabayashi, K.; Fujita, M.; Ajiki, H.; Sigrist, M. {\it Phys. Rev. B} {\bf 1999}, 59, 8271Ð8282. 
\item Peres, N. M. R.; Neto, A. H. C.; Guinea, F. {\it Phys. Rev. B} {\bf 2006}, 73, Art. No. 195411.
\item Ouyang, Y.; Yoon, Y.; Fodor, J. K.; Guo, J. {\it Appl. Phys. Lett.} {\bf 2006}, 89, Art. No. 203107.
\item Barone, V.; Hod, O.; Scuseria, G. E. {\it Nano Lett.} {\bf 2006}, 6, 2748Ð2754.   
\item Areshkin, D. A.; Gunlycke, D.; White, C. T. {\it Nano Lett.} {\bf 2007}, 7, 204Ð210.   
\item Hod, O.; Barone, V.; Peralta, J. E.; Scuseria, G. E. {\it Nano Lett.} {\bf 2007}, 7, 2295Ð2299.   
\item Obradovic, B.; Kotlyar, R.; Heinz, F.; Matagne, P.; Rakshit, T.; Giles, M. D.; Stettler, M. A.; Nikonov, D. E. {\it Appl. Phys. Lett.} {\bf 2006}, 88, 142102. 
\item Chen, Z.; Lin, Y.-M.; Rooks, M. J.; Avouris, P. {\it Physica E} {\bf 2007}, 40, 228Ð232. 
\item Han, M. Y.; Ozyilmaz, B.; Zhang, Y.; Kim, P {\it Phys. Rev. Lett.} {\bf 2007}, 98, Art. No. 206805.
\item Basu, D.; Gilbert, M. J.; Register, L. F.; MacDonald, A. H.; Banerjee, S. K. {\it Appl. Phys. Lett.} {\bf 2007}, 92, 042114. 
\item Li, T. C.; Lu, S.-P. {\it Phys. Rev. B} {\bf 2008}, 77, 085408. 
\item Tomita, A.; Tamai, Y. {\it J. Phys. Chem.} {\bf 1974}, 78, 2254Ð2258. 
\item Novoselov, K. S.; Geim, A. K.; Morozov, S. V.; Jiang, D.; Zhang, Y.; Dubonos, S. V.; Grigorieva, I. V.; Firsov, A. A. {\it Science} {\bf 2004}, 306, 666Ð669.  
\item Klinke, C.; Bonard, J.-M.; Kern, K. {\it J. Phys. Chem. B} {\bf 2004}, 108, 11357Ð11360.  
\item Ishigami, M.; Chen, J. H.; Cullen, W. G.; Fuhrer, M. S.; Williams, E. D. {\it Nano Lett.} {\bf 2007}, 7, 1643.   
\item Methane formation in the reaction given by eq 1 could occur either by dissociation and dissolution of molecular hydrogen in the Fe nanoparticle and subsequent hydrogen reaction with carbon atoms at the metal-graphene interface or, as is more likely, by dissolution of carbon atoms from the metal-graphene interface through the Fe nanoparticles and subsequent reaction with hydrogen at the Fe surface.
\item de Gennes, P. G.; Brochard-Wyart, F.; Quere, D. {\it Capillary and Wetting Phenomena -- Drops, Bubbles, Pearls, Waves}; Springer: New York, {\bf 2002}.
\item Thomas, J. M., {\it Chemistry and Physics of Carbon}; Marcel Dekker: New York, {\bf 1965}; Vol. 1. 
\end{enumerate}

\newpage

\begin{figure}
\begin{center}
\includegraphics[width=4.5in]{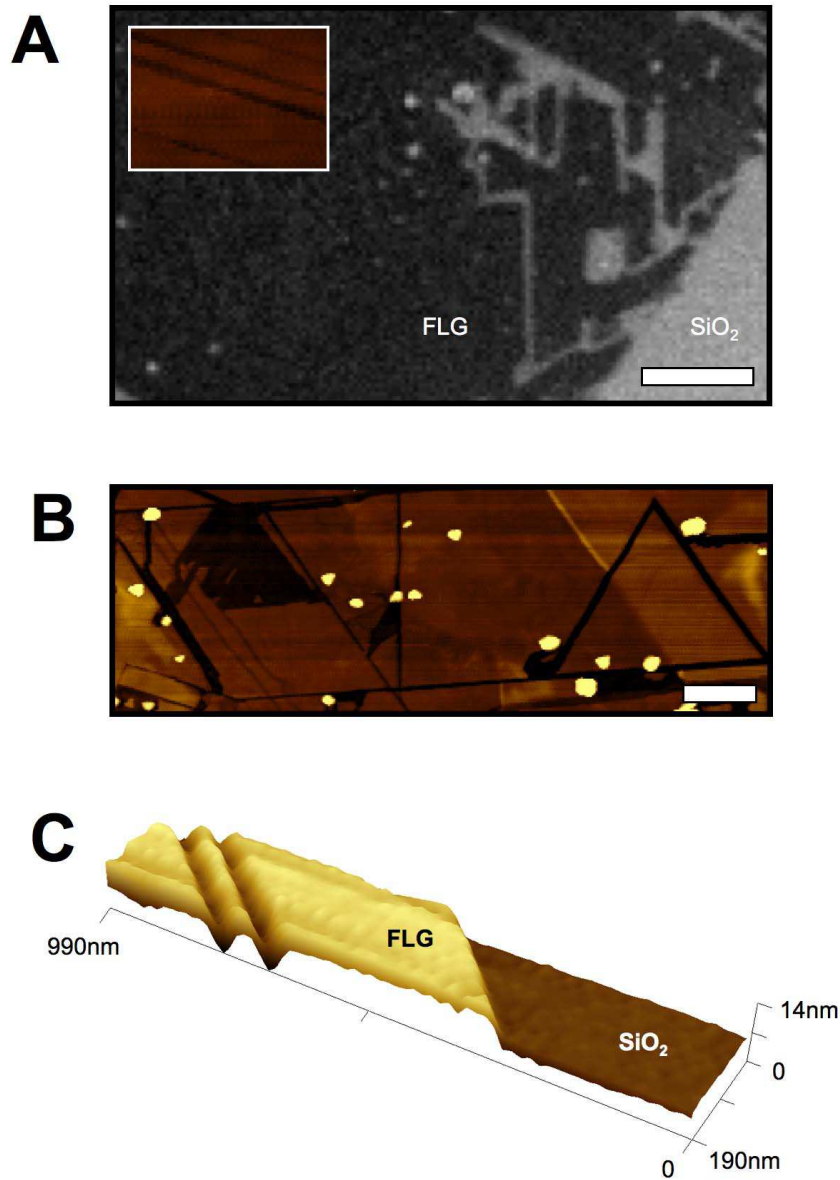}
\caption{Single-particle crystallographic etching of few-layer graphene. (A) SEM image of nanotrenches etched in FLG (dark) supported on SiO$_{2}$/Si substrate (light); scale bar is 800 nm. Small circles are Fe nanoparticles. Inset image, width 315 nm, shows an AFM image (color scale 15 nm) of a graphene nanoribbon defined using parallel etched trenches, of measured width $\sim$35 nm. (B) AFM image (color scale 15 nm) of etched FLG sample, showing straight nanotrenches etched down to the substrate defining a triangle of side length $\sim$650 nm; scale bar is 250 nm. (C) 3D AFM surface plot of etched FLG sample, showing parallel straight nanotrenches.}
\end{center}
\end{figure}
\newpage

\begin{figure}
\begin{center}
\includegraphics[width=5in]{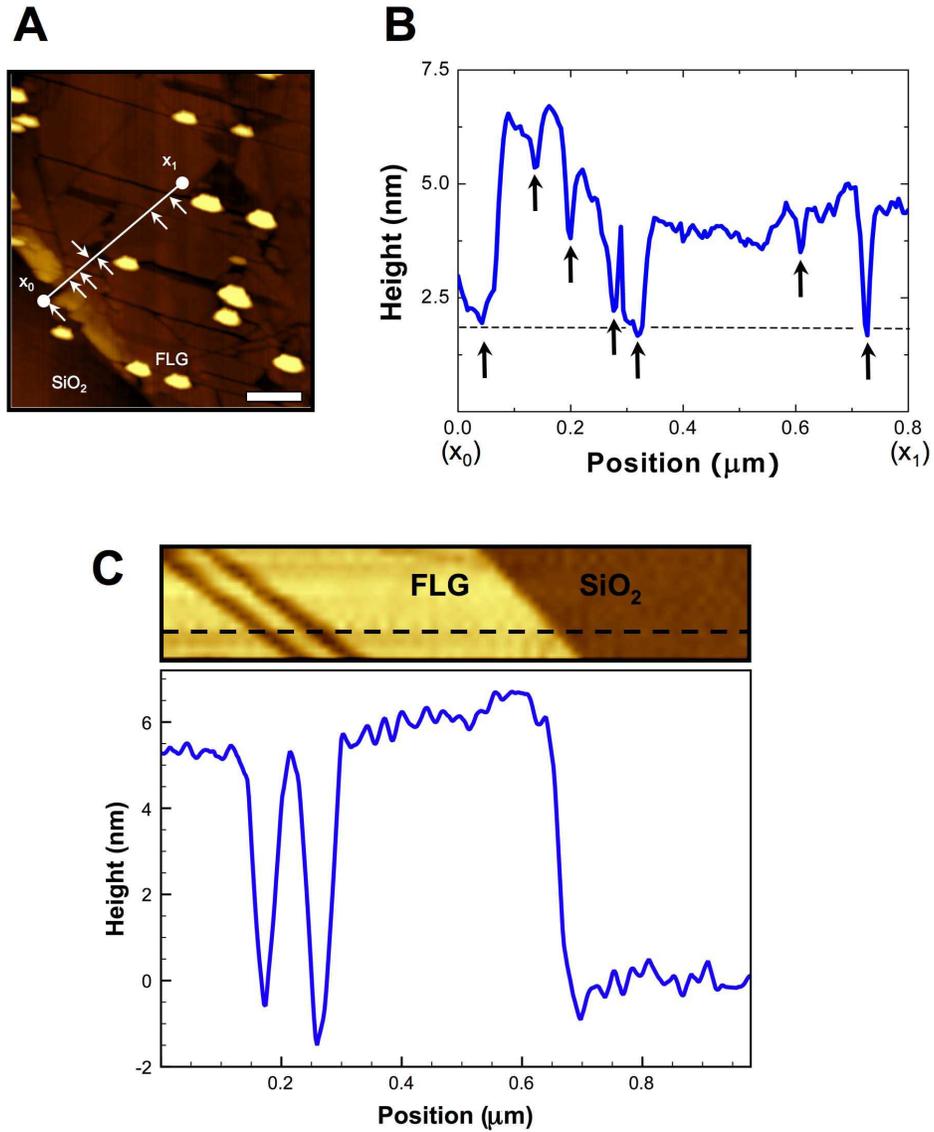}
\caption{(A) AFM image of etched FLG sample; color scale is 14 nm. (B) Height profile along the line in (A), showing nanotrenches etched down to the underlying SiO$_{2}$/Si substrate: leftmost arrow indicates substrate, while other arrows indicate etched trenches. Because of the finite size of the AFM tip, measured nanotrench heights (for trenches of width $<$ 20 nm) appear shorter. The height increase near the edge of the film is due to a fold in the FLG. (C) AFM image (color scale 14 nm) of etched FLG shown in Figure 1C with parallel nanotrenches etched by catalyst nanoparticles, with height profile along the dotted line showing nanotrenches etched down to the underlying SiO$_{2}$/Si substrate.}
\end{center}
\end{figure}

\newpage

\begin{figure}
\begin{center}
\includegraphics[width=5in]{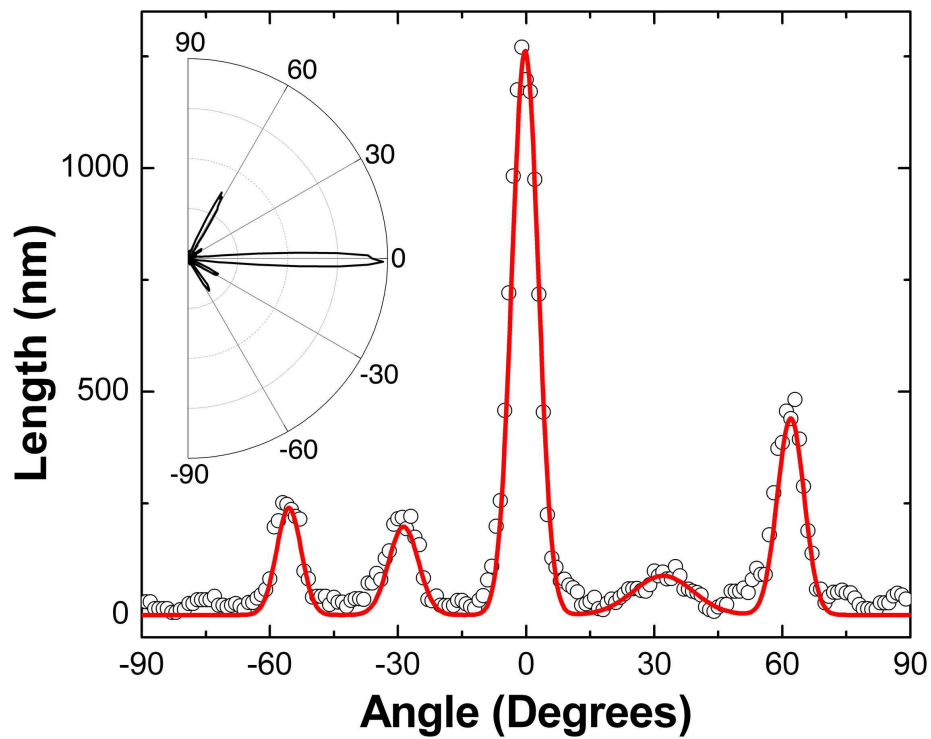}
\caption{Histogram for AFM image of Figure 2A showing total etched nanotrench length versus trench angle, chosen so the main peak is at 0$^\circ$ (FLG edge is at $\sim$34$^\circ$). Open circles show the data averaged over seven adjacent points; red curve is a five-peak Gaussian fit. Inset shows histogram data (open circles) as a polar plot.}
\end{center}
\end{figure}
\newpage

\begin{figure}
\begin{center}
\includegraphics[width=3in]{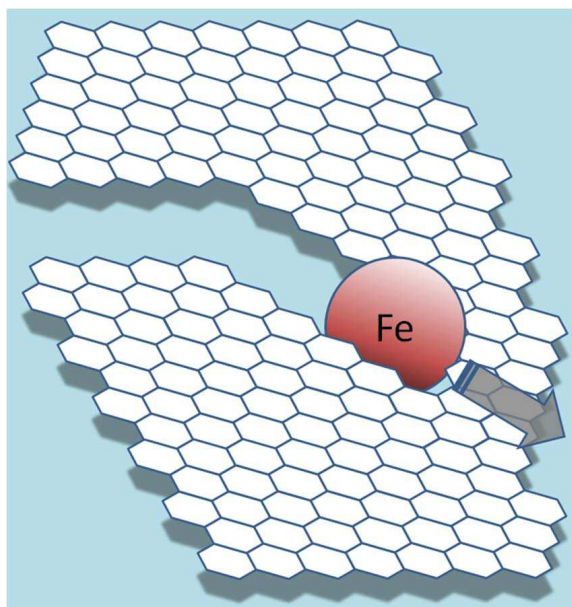}
\caption{Schematic of crystallographic few-layer graphene (FLG) etching process. Thermally activated Fe nanoparticles react with active carbon atoms at graphene edges, moving and breaking FLG carbon-carbon bonds along specific crystallographic directions.}
\end{center}
\end{figure}

	\end{document}